\documentclass[prc,preprint,showpacs,tightenlines,%
nofootinbib]{revtex4}

\usepackage{graphicx}  
\usepackage{bm}  
\usepackage{amsmath}
\usepackage{epsf}
\newcommand{\beq}{\begin{equation}}
\newcommand{\eeq}{\end{equation}}
\newcommand{\bea}{\begin{eqnarray}}
\newcommand{\eea}{\end{eqnarray}}

\newcommand{\simgt}{\stackrel{>}{{}_\sim}}

\newcommand{\textfrac}[2]{{\textstyle\frac{#1}{#2}}}

\begin{document}
\preprint{\tiny{FZJ-IKP(TH)-2004-03}}
\preprint{\tiny{HISKP-TH-04/05}}
\preprint{\tiny{INT-PUB 04-08}}
\title{The Four-Boson System with Short-Range Interactions}
\author{L. Platter}\email{l.platter@fz-juelich.de}
\affiliation{Forschungszentrum J{\"u}lich, Institut f{\"ur} Kernphysik,
    D-52425 J{\"u}lich, Germany}
\affiliation{Helmholtz-Institut f\"ur Strahlen- und Kernphysik (Theorie),
Universit\"at Bonn, Nu\ss allee 14-16, D-53115 Bonn, Germany}
\author{H.-W. Hammer}\email{hammer@phys.washington.edu}
\affiliation{Institute for Nuclear Theory, University of Washington, Seattle, 
WA 98195, USA}
\author{Ulf-G. Mei{\ss}ner}\email{meissner@itkp.uni-bonn.de}
\affiliation{Helmholtz-Institut f\"ur Strahlen- und Kernphysik (Theorie),
Universit\"at Bonn, Nu\ss allee 14-16, D-53115 Bonn, Germany}
\affiliation{Forschungszentrum J{\"u}lich, Institut f{\"ur} Kernphysik,
    D-52425 J{\"u}lich, Germany}
\date{June, 2004}
\begin{abstract}
We consider the non-relativistic four-boson system with short-range forces
and large scattering length in an effective quantum mechanics approach. 
We construct the effective interaction potential at leading order in the
large scattering length and compute the four-body binding energies 
using the Yakubovsky equations. Cutoff independence of the four-body binding 
energies does not require the introduction of a four-body force.
This suggests that two- and three-body interactions are sufficient 
to renormalize the four-body system. We apply the equations to $^4$He
atoms and calculate the binding energy of the $^4$He tetramer. We observe
a correlation between the trimer and tetramer
binding energies similar to the Tjon line in nuclear physics.
Over the range of binding energies relevant to $^4$He atoms, the 
correlation is approximately linear.
\end{abstract}
\maketitle
\section{Introduction}
\label{sec:intro}
Effective theories are ideally suited to describe the low-energy
properties of physical systems in a model-independent way. 
They can be applied to any system that has a separation of scales,
which can be a fundamental property of the underlying theory
or simply a kinematical suppression.
The long-distance degrees of freedom must be included dynamically 
in the effective theory, while short-distance physics enters only 
through the values of a few coupling constants, often called
low-energy constants.
Effective theories are widely used in many areas of physics.  
Recently, a considerable effort was devoted to applying
effective field theories in nuclear and atomic physics. For overviews
of these programs, see e.g. Refs.~\cite{Beane:2000fx,Bedaque:2002mn}
and \cite{Hammer:2003ih,Andersen:2003qj}.
If there is no exchange of massless particles, any interaction will
appear short ranged at sufficiently low energy. One can then use
a very general effective theory with short-range interactions only
to describe the universal low-energy properties of the system. 
Such a theory can be applied to a wide range of systems 
from nuclear and particle physics to atomic and molecular physics.

Most previous work in this area was done using a non-relativistic 
effective field theory (EFT) with contact interactions. Particularly
interesting are few-boson systems with large scattering length.
They are characterized by an unnaturally large two-body scattering length $a$
which is much larger than the typical low-energy length scale $l$ given
by the range of the interaction. Such systems
display a number of interesting effects and universal properties that are 
independent of the details of the interaction at short distances of order $l$. 
If $a>0$, e.g., there is a shallow two-body bound state with binding energy 
$B_2=\hbar^2/(Ma^2)+{\cal O}(l/a)$, where $M$ is the mass of the particles. 
Low-energy observables can generally be described in a controlled expansion 
in $l/|a|$. In the two-boson system, the effective 
theory reproduces the effective range expansion 
(cf.~Refs.~\cite{Kaplan:1998tg,vanKolck:1998bw}) but
the structure of the three-boson system with large scattering length
is richer. In Refs.~\cite{Bedaque:1998kg,Bedaque:1998km}, 
it was found that both two- and three-body contact interactions
are required at leading order for the consistent renormalization of the 
three-body system. Interestingly, the renormalization group behavior of the 
three-body interaction is governed by an ultraviolet limit cycle.
This implies that at leading order in $l/|a|$, the properties of the 
three-boson system with large scattering length are not determined by 
two-body data alone and one piece of three-body information 
(such as a three-body binding energy) 
is required as well. In the EFT, this  information can 
conveniently be parameterized by the three-body parameter $\Lambda_*$ 
introduced in Refs.~\cite{Bedaque:1998kg,Bedaque:1998km}.
These general findings confirm and extend previous work by Efimov who 
derived many general features of the three-body problem with large
scattering length \cite{Efimov71,Efimov79}.

While the EFT formulation has been very successful, it is not
the only possible formulation of an effective theory for this problem. 
Lepage has advocated the framework of non-relativistic quantum 
mechanics with an effective interaction potential \cite{Lepage:1997cs}.
The contact operators in the field theory are the replaced by 
an ``effective potential'' built from smeared out
$\delta$-function potentials and derivatives thereof. 
In the case $a>0$,
this approach has been applied to the three-boson system  
by Wilson \cite{Wilson00} and Mohr \cite{Mohr:2003du}. They
confirmed the results of Refs.~\cite{Bedaque:1998kg,Bedaque:1998km}
and were able to calculate the binding energies for the three-boson 
system to extremely high accuracy. 

In this paper, we consider the four-body system with short-range 
interactions. The four-body problem has previously been studied in 
variety of approaches. Early studies include the Yakubovky equations
for local potentials using the Hilbert-Schmidt expansion \cite{Tjo75},
the Schr\"odinger equation with separable two-body potentials \cite{Gibs76},
and field-theoretical models with
separable expansions of the three-body T-matrix \cite{Fons76}.
The four-body problem of $^4$He atoms was investigated by Nakaichi-Maeda
and Lim using the  Yakubovsky equations with a unitary pole approximation 
for the S-wave (2)+(2) and (3)+(1) subamplitudes \cite{Naka83}. 
For a review of these and other early studies see, e.g., 
Refs.~\cite{Tjon78,Lim84}. For an overview of recent calculations for 
the four-body system of $^4$He atoms, see Refs.~\cite{Blume:2000,Fili02}.
A general review of theoretical studies of small $^4$He clusters
can be found in Ref.~\cite{Whal94}. In Ref.~\cite{Kama01},
a benchmark calculation comparing various modern calculational approaches
to the nuclear four-body problem was carried out.

The purpose of this paper is to study the four-boson system with short-range 
interactions and large scattering length in an effective theory.
We will work at leading order in $l/|a|$ and use the framework of 
non-relativistic quantum mechanics to construct an effective interaction 
potential. This approach has the advantage that one can immediately 
start from the well-known Yakubovsky equations for the four-body 
system \cite{Yakubovsky:1966ue}. The four-boson binding energies are 
obtained by solving the Yakubovsky equations for the effective 
interaction potential.
The solution of the four-boson problem in effective theory 
is important in several respects:

First, it can immediately be applied to the atomic problem of $^4$He 
atoms and is a first step towards the four-body problem in nuclear
physics which is complicated by spin and isospin.
The scattering length of $^4$He atoms $a\approx
100$ \AA\ is much larger than its effective range $r_e 
\approx 7$ \AA\ which can be taken as an estimate of the
natural low-energy length scale $l$. $^4$He atoms are therefore an 
ideal application for our theory.  
The three-body system of $^4$He atoms  has been investigated in 
Refs.~\cite{Bedaque:1998km,Braaten:2002jv,Braaten:2002sr} using
effective field theory. While the universal properties of the three-body 
system of $^4$He atoms were discussed in Ref.~\cite{Braaten:2002jv},
this has not been done for the four-body system.

Second, the renormalization of the four-body system in an effective
theory is an open question. It is clear that low-energy four-body
observables must depend on a two-body parameter and a three-body parameter. 
However, it is not known whether a four-body parameter is also 
required to calculate low-energy four-body observables up to 
corrections suppressed by $l/|a|$.

The theoretical situation concerning this question appears
confusing. On the one 
hand, there is a renormalization argument for $\delta$-function pair
potentials that indicates that a new four-body parameter is required
to calculate four-body binding energies \cite{AFG95}. On the other
hand, Amado and Greenwood have evaluated the trace of the four-body
kernel and concluded that the Efimov effect is absent in 
the four-body system \cite{Amado:mf}. This result suggests
that a four-body parameter should not be necessary at leading
order in $l/|a|$.  There is some circumstantial evidence in favor of 
the latter possibility from the four-body
problem in nuclear physics. There is a correlation called the 
\lq\lq Tjon line'' between the binding energy $B_t$ of the triton
and the binding energy $B_\alpha$ of the $\alpha$  particle \cite{Tjo75}.
Calculations of these binding energies using modern 
phenomenological nucleon-nucleon
interaction potentials give results that underestimate both 
binding energies but cluster along a line in the 
$B_t$-$B_\alpha$ plane.
By adding a three-body potential whose strength is adjusted to get
the correct value for $B_t$, one also gets an accurate result for 
$B_\alpha$ (cf. Ref.~\cite{Nogga:2000uu}). This conclusion also holds 
for chiral nuclear potentials derived from an effective field theory
with explicit pions \cite{Epelbaum:2002vt}.
The aim of the present work is to study the question about
the requirement of the four-body interaction by an explicit calculation
in the controlled environment of an effective theory.

The paper is organized as follows. In Sec.~\ref{sec:effT},
we will review the description of the 
two- and three-body bound state problem in effective theory
and extend this framework to the four-body bound state problem. 
In Sec.~\ref{sec:renres}, we will discuss the renormalization and 
present numerical results for the case of $^4$He atoms. Finally, we close 
with a summary and outlook in Sec.~\ref{sec:conc}.

\section{Few-Body Bound State Equations in Effective Theory}
\label{sec:effT}
The effective low-energy interaction potential generated by a 
non-relativistic EFT with short-range interactions can be written
down in a momentum expansion. In the two-body S-wave sector,
it takes the general form
\beq
\langle {\bf k'} | V | {\bf k} \rangle =
\lambda_2 + \lambda_{2,2} (k^2+k'^2)/2 +\ldots\,,
\label{effpot}
\eeq
where ${\bf k}$ and ${\bf k'}$ are the relative three-momenta of
the incoming and outgoing particles, respectively.
Because of Galilean invariance, the interaction can only depend on the 
relative momenta.
Similar expressions can be derived for three- and higher-body 
interactions. The exact form of the potential depends on the specific
regularization scheme used. The low-energy observables, however, 
are independent of the regularization scheme (up to higher order corrections)
and one can choose a convenient scheme for practical calculations.\footnote{
For a comparison of different regularization schemes in the nuclear
two-body problem and chiral perturbation theory, see Refs.~\cite{Steele:1998un}
and \cite{Espriu:1993if}, respectively.}  

In a momentum cutoff scheme,
the potential in Eq.~(\ref{effpot}) can be regularized by multiplying with 
a Gaussian regulator function, $\exp[-(k^2+k'^2)/\Lambda^2]$, with
the cutoff parameter $\Lambda$. This factor strongly suppresses 
high-momentum modes in the region $k,k' \simgt \Lambda$ where the 
effective potential is not valid. 
The cutoff dependence of the coefficients $\lambda_2(\Lambda)$, 
$\lambda_{2,2}(\Lambda)$, $\ldots$ is determined by the requirement
that low-energy observables are independent of $\Lambda$. Of course,
the expansion in Eq.~(\ref{effpot}) is only useful in conjunction with a 
power counting scheme that determines the relative importance of the various
terms at low energy. In the case of large scattering length $a$, 
the leading order is given by the $\lambda_2$ term which must 
be iterated to all orders, while the other terms give rise to higher-order
corrections that can be included perturbatively  
\cite{Kaplan:1998tg, vanKolck:1998bw}.
In this paper, we will work to leading order in the large scattering length
$a$ and include only the $\lambda_2$ term.

In the three-body system, 
a momentum-independent three-body interaction term $\lambda_3$ must be
included together with $\lambda_2$ already at leading order 
\cite{Bedaque:1998kg,Bedaque:1998km}. Without this three-body interaction
low-energy observables show a strong cutoff dependence and
the system can not be renormalized. Effective range effects
and other higher-order corrections can be included as well
\cite{Hammer:2001gh,Bedaque:2002yg,Afnan:2003bs}.

The power counting for the four-body system has not been formulated 
yet. In order to see whether the minimal set of interactions 
$\lambda_2$ and $\lambda_3$ is sufficient, we will 
calculate the bound states in the four-body system and study
their cutoff dependence. We will use a momentum cutoff regularization
scheme as described above. A strong cutoff dependence of the binding
energies would indicate that a four-body interaction term is required.
If the four-body binding energies are stable under variations of the 
cutoff, this would suggest that the four-body interaction is a 
subleading effect.

In order to set up our conventions and formalism,
we will first review the bound state equations for the two- and 
three-body system and then set up the four-body equations.

\subsection{The Two-Body Sector}
\label{sec:effT:2body}
We write the leading order two-body effective potential in momentum space as:
\begin{equation}
\langle {\bf p}|V|{\bf q}\rangle=
\langle {\bf p}|g \rangle \lambda_2 \langle g|{\bf q} \rangle ~,
\label{effpot_2}
\end{equation}
where $\lambda_2$ denotes the two-body coupling constant and
${\bf q}$ (${\bf p}$) are the relative three-momenta in the incoming
(outgoing) channel. The regulator functions
\beq
\langle {\bf p}|g\rangle \equiv g(p)=\exp(-p^2/\Lambda^2)~,
\eeq
suppress the contribution from high momentum states. In the few-body 
literature, they are often called ``form factors''. 

Our normalization for plane wave and spherical wave states is 
\bea
\langle {\bf p}|{\bf p'}\rangle&=&\delta^{(3)}({\bf p}-{\bf p'})~,
\qquad
\langle p\;l\; m| p'\;l'\; m' \rangle = \frac{\delta(p-p')}{pp'}
\delta_{ll'}\delta_{mm'}~,
\eea
and the S-wave projection of the plane wave state with momentum ${\bf p}$ is
\beq
\langle p\; 0\; 0 |{\bf p'}\rangle\equiv \langle p|{\bf p'}\rangle
=\frac{1}{(4\pi)^{1/2}}\frac{\delta(p-p')}{p p'}~.
\eeq
For convenience, we will work in units where the mass $M$ of the 
bosons and Planck's constant $\hbar$ are set to unity: $M=\hbar=1$.

The interaction (\ref{effpot_2}) is separable and the Lippmann-Schwinger
equation for the two-body problem can be solved analytically.
The two-body t-matrix can be written as \cite{Ziegelmann}:
\begin{equation}
\label{t-matrix}
t(E)=|g\rangle\tau(E)\langle g|~,
\end{equation}
where $E$ denotes the total energy. The two-body propagator $\tau(E)$ is
then given by
\begin{equation}
\tau(E)=\left[1/\lambda_2-4\pi\int_0^\infty\hbox{d}q\,
  q^2\frac{g(q)^2}{E-q^2}\right]^{-1}~.
\end{equation}

A two-body bound-state appears as a simple pole in the two-body propagator
$\tau$ at energy $E=-B_2$. Thus the two-body coupling constant 
$\lambda_2(B_2,\Lambda)$ can be fixed from the two-body binding energy $B_2$,
which is directly related to the scattering length by $a=1/\sqrt{B_2}$
at leading order in $l/|a|$.
The integrals appearing in the propagator can be expressed
through the complementary error function
\beq
\hbox{erfc}(x)=1-\frac{2}{\sqrt{\pi}}\int_0^x e^{-t^2} \hbox{d} t~,
\eeq
and for $E<0$, we obtain for the inverse propagator:
\beq
\tau(E)^{-1}=2\pi^2 \left[ 
\sqrt{B_2} \exp\left(\frac{2B_2}{\Lambda^2}\right)
\hbox{erfc}\left(\frac{\sqrt{2 B_2}}{\Lambda}\right)
-\sqrt{-E} \exp\left(\frac{-2E}{\Lambda^2}\right)
\hbox{erfc} \left(\frac{\sqrt{-2 E}}{\Lambda} \right)\right]~.
\label{eq:tau2}
\eeq
A similar expression for the propagator can be obtained for positive 
energies by adding a small imaginary part to $E$, but will not be 
required for our purposes.
\subsection{The Three-Body Sector}
\label{sec:effT:3body}
The low-energy properties of the three-body system for a given
effective potential can be obtained by solving the Faddeev equations
\cite{Faddeev:1960su}. 
Faddeev's idea was to decompose the full three-body wave function $\Psi$
into so-called ``Faddeev components'' in order to avoid the problem of 
disconnected contributions in the three-body scattering problem.
For the three-body problem with two- and three-body interactions, the
full wave function can be decomposed into four components \cite{Meier:hj}:
one for each two-body subcluster and one for the three-body 
cluster.\footnote{Note, however, that other
decompositions involving only three Faddeev components are possible as
well.}
For identical bosons, the three-body wave function is fully symmetric under 
exchange of particles and the Faddeev equations simplify considerably.
In this case, one only needs to solve equations involving  one of the 
two-body Faddeev components and the three-body component.
The two remaining two-body components can be obtained by permutations 
of particles. For more details on the Faddeev equations, we refer 
the reader to the literature \cite{Gloeckle,Ziegelmann}.

We follow Gl\"ockle and Meier \cite{Meier:hj} and decompose
the full three-body wave function as
\beq
\Psi=(1+P)\psi+\psi_{3}~,\qquad \mbox{where}\quad
 P=P_{13}P_{23}+P_{12}P_{23}
\label{eq:defP}
\eeq
is a permutation operator that generates the two not explicitly included
Faddeev components from $\psi$. The operator $P_{ij}$ simply
permutes particles $i$ and $j$. The Faddeev equations for $\psi$ and
$\psi_3$ in operator form are then
\bea
\psi &=& G_0\, t\, P\, \psi+G_0\, t\, \psi_{3}~,\nonumber \\
\psi_{3} &=& G_0\, t_{3}\, (1+P)\, \psi~,
\label{eq:faddeev}
\eea
where $G_0$ denotes the free three-particle propagator.
$t$ is the two-body t-matrix for the two-body subsystem 
described by the component $\psi$.
$t_3$ is the three-body t-matrix defined by the solution of the 
three-body Lippmann-Schwinger equation with the leading order 
three-body effective interaction
\beq
V_{3}=|\xi\rangle\lambda_3\langle\xi|~,
\label{def:3pot}
\eeq
only. Since $V_3$ is separable, we can solve for $t_{3}$ exactly and obtain
\beq
t_{3}(E)=|\xi\rangle\tau_{3}(E)\langle\xi|~,\qquad
\mbox{where} \quad
\tau_{3}(E)=\left[\frac{1}{\lambda_3}-\langle\xi|G_0|\xi\rangle\right]^{-1}~.
\eeq
The three-body regulator function $|\xi\rangle$ will be
specified later. Note that $t_3$ is only a technical construct that is
generally cutoff dependent and not observable. The physical three-body
t-matrix always includes both two- and three-body forces.
Since we are interested only in the binding energies and not in the
wave functions, we can eliminate the component $\psi_3$ and obtain
\beq 
\psi=G_0\, t\, P\, \psi +G_0\, t\, G_0\, t_{3}\, (1+P)\,\psi~.
\label{faddeq}
\eeq
The component $\psi_3$ can easily be recovered by using
the second line of Eq.~(\ref{eq:faddeev}).

We now derive an explicit representation of Eq.~(\ref{faddeq})
in momentum space. We will illustrate this procedure by showing
some details for the first term on the right-hand side
of Eq.~(\ref{faddeq}). The extension to include the second term
containing three-body interaction is straightforward
and we will only quote the final result. 
Furthermore, in order to understand the renormalization of the three-body 
problem it is instructive to consider the case without a three-body force 
first. The natural Jacobi momenta 
are given by:
\bea
\label{eq:jacobi:3body}
{\bf u_1}=\frac{1}{2}({\bf k_1}-{\bf k_2})~,\qquad
{\bf u_2}=\frac{2}{3}\left[{\bf k_3}-\frac{1}{2}({\bf k_1}+{\bf k_2})\right]~.
\eea
Here and in the following sections, we will only take $S$-waves
into account. Hence, we can project all operators accordingly
and define the $S$-wave projection operator:
\beq
\int \hbox{d}u_1 \:u_1^2\:\hbox{d}u_2 \:u_2^2 \:|u_1 u_2\rangle\langle u_1
u_2|\equiv \int \mathcal{D}u\: u^2\:|u_1 u_2\rangle\langle u_1 u_2|~.
\label{eq:S-proj}
\eeq
Using the definition $\langle u_1 u_2|\psi\rangle \equiv \psi(u_1,u_2)$,
we can write the Faddeev equation in momentum space as:
\bea
\nonumber
\psi(u_1,u_2)&=&\langle u_1 u_2|G_0 t P|\psi\rangle=4\pi G_0(u_1,u_2)g(u_1)
\int\mathcal{D}u^\prime{u^\prime}^2
\int\mathcal{D}u^{\prime\prime}{u^{\prime\prime}}^2
g(u_1^\prime) \tau(E-\textfrac{3}{4}{u_2^\prime}^2)
\\ &&
\qquad\quad\times
\frac{\delta(u_2-u_2^\prime)}
{{u_2^\prime}^2}\langle{u_1}^\prime{u_2}^\prime|P
|{u_1}^{\prime\prime}{u_2}^{\prime\prime}\rangle
\langle{u_1}^{\prime\prime}{u_2}^{\prime\prime}|\psi\rangle~,
\label{eq:fad0}
\eea
where the factor of $4\pi$ arises from our normalization of the S-wave 
projected two-body t-matrix. The free propagator for three particles 
in their center of mass is given by
\beq
G_0(u_1,u_2)=\left[E-E_{kin}\right]^{-1}
 =\left[E-u_1^2-\textfrac{3}{4} u_2^2 \right]^{-1}~,
\eeq 
where
$E_{kin}$ denotes the kinetic energy.
Furthermore, it should be noted that the two-body propagator $\tau$
in Eq.~(\ref{eq:fad0}) is evaluated at the energy in the corresponding
two-body subsystem: $E-\textfrac{3}{4}{u_2^\prime}^2$.
The permutation operator $P$ can be written as:
\beq
\langle u_1 u_2| P | u_1^\prime u_2^\prime\rangle=\int_{-1}^{1}\hbox{d} x
\,\frac{\delta(u_1 -\pi(u_2,u'_2))
\delta(u_1^\prime-\pi(u'_2,u_2))}{\pi(u_2,u'_2)^2\:
\pi(u'_2,u_2)^2}~,
\eeq
\beq
\pi(u_2,u'_2)=\sqrt{\textfrac{1}{4}u_2^2+{u_2^\prime}^2
+u_2 u_2^\prime x}.
\label{eq:defpi}
\eeq
Using this representation of $P$, we can write the integral equation as:
\bea
\psi(u_1,u_2) &=& 4\pi\, G_0(u_1,u_2)\, g(u_1)\, 
  \tau(E-\textfrac{3}{4} u_2^2) \int_0^\infty 
  \hbox{d} u'_2 {u'_2}^2 
  \int_{-1}^1\hbox{d} x\, g(\pi(u_2,u'_2))
  \nonumber \\ && \qquad \times
  \psi(\pi(u'_2,u_2), u'_2)~.
\eea
This is an homogeneous integral equation in two variables.
It can be further simplified by defining a new function $F(u_2)$ of only
one variable via
\beq
\psi(u_1, u_2)=G_0(u_1,u_2)\,g(u_1)\,\tau(E-{\textstyle\frac{3}{4}}u_2^2)
F(u_2)~,
\eeq
leading to the integral equation
\bea
\label{faddeev1}
F(u_2)&=&4\pi\int_0^\infty\hbox{d}u'_2 {u'_2}^2
\int_{-1}^1 \hbox{d} x \: g(\pi(u_2,u'_2))
G_0(\pi(u'_2,u_2),u'_2) g(\pi(u'_2,u_2))
\nonumber\\
&&\qquad\qquad\qquad\qquad\times
\tau(E-\textfrac{3}{4} {u'_2}^2) F(u'_2)~.
\label{eq:faddeev2}
\eea

The three-body binding energies are given by those values of $E$
for which Eq.~(\ref{eq:faddeev3}) has a nontrivial solution.
By expressing the two-body coupling constant $\lambda_2$ in terms
of the binding energy of the shallow two-body bound state in
Eq.~(\ref{eq:tau2}), we have already renormalized the two-body problem.

\begin{figure}[tb]
\centerline{\includegraphics*[width=5.in,angle=0]{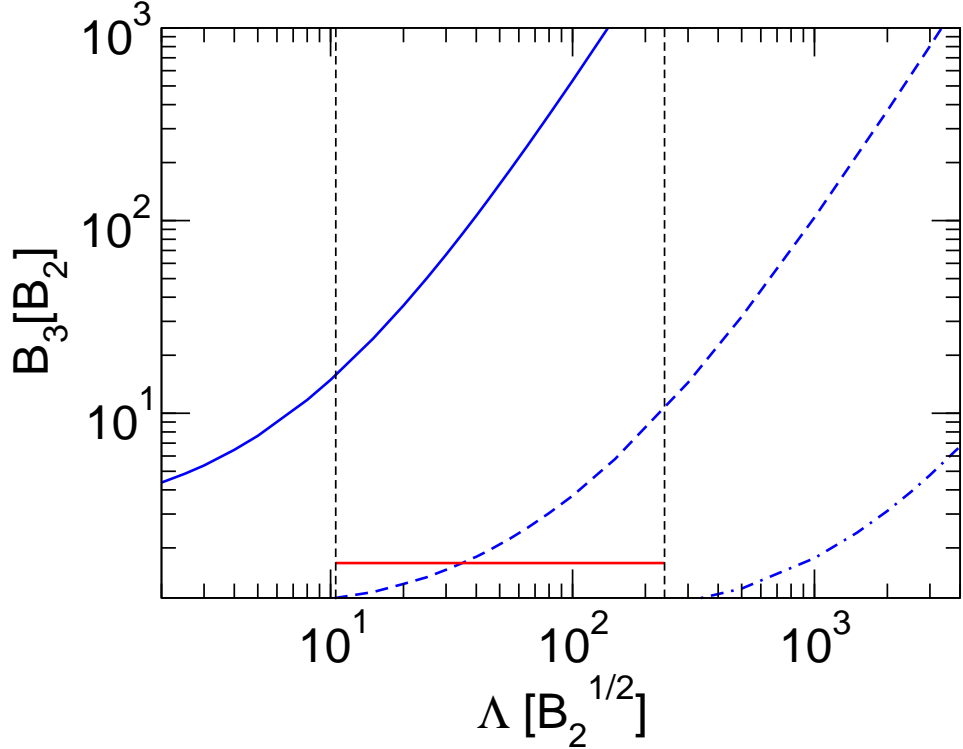}}
\caption{\label{fig:lambdab3}
The shallowest three-body binding energies indicated by
the solid, dashed, and dash-dotted lines as a function of the 
momentum cutoff $\Lambda$. 
The vertical dashed line indicates the cutoff range in which the
three-body system has exactly two bound states. The horizontal
solid line shows the energy at which the shallowest three-body state
is fixed.
}
\end{figure}

The three-body system is stabilized against the Thomas collapse by 
the presence of the momentum cutoff $\Lambda$. 
No three-body force is required for this purpose. 
After the cutoff is introduced, 
there are no bound states with binding energies 
$B_3 \simgt \Lambda^2$.\footnote{This momentum cutoff $\Lambda$ can
also be thought of as introducing a finite range $r\sim 1/\Lambda$
for the pair potential in position space. The Thomas collapse is then 
stopped when the bound state size is of the same order as 
the range of the potential.} 
However, the three-body binding energies
depend strongly on the value of the cutoff $\Lambda$.
This is illustrated in Fig.~\ref{fig:lambdab3} where
the shallowest three-body binding energies are indicated by
the solid, dashed, and dash-dotted lines.
The three-body system has exactly two bound states in the cutoff range
indicated by the two vertical dashed lines.

We now include the second term on the right-hand side of Eq.~(\ref{faddeq})
which contains the three-body force. The derivation of an explicit 
representation in momentum space proceeds as for the first term.
The full equation including both terms then reads
\bea
F(u_2)&=&4\pi\int_0^\infty\hbox{d}u'_2 {u'_2}^2 
  \int_{-1}^1 \hbox{d} x \: g(\pi(u_2,u'_2))
G_0(\pi(u'_2,u_2),u'_2) g(\pi(u'_2,u_2))
\nonumber\\
&&\qquad\qquad\qquad\qquad\times
\tau(E-\textfrac{3}{4} {u'_2}^2) F(u'_2)
\nonumber\\[0.4cm]
&+&(4\pi)^3 \int_0^\infty\hbox{d}u_1^\prime {u_1^\prime}^2 g(u_1^\prime)
 G_0(u_1^\prime,u_2)\tau_3(E)\xi(u_1^\prime,u_2)
\nonumber\\
&&\times\Bigg[\int_0^\infty\hbox{d}u_1^{\prime\prime}{u_1^{\prime\prime}}^2
\int_0^\infty \hbox{d}u_2^{\prime\prime}{u_2^{\prime\prime}}^2
\,\xi(u_1^{\prime\prime},u_2^{\prime\prime}) g(u_1^{\prime\prime})
G_0(u_1^{\prime\prime},u_2^{\prime\prime})
\tau(E-\textfrac{3}{4}{u_2^{\prime\prime}}^2)
 F(u_2^{\prime\prime})\nonumber\\[0.4cm]
&&\quad+\int_0^\infty\hbox{d}u_2^{\prime\prime}{u_2^{\prime\prime}}^2
\int_0^\infty\hbox{d}u_2^{\prime\prime\prime}{u_2^{\prime\prime\prime}}^2
\int_{-1}^1\hbox{d}x \, \xi(\pi(u''_2,u'''_2),u''_2)
g(\pi(u'''_2,u''_2))
\nonumber\\
&&\qquad\qquad\qquad\qquad\qquad\qquad\times 
G_0(\pi(u'''_2,u''_2),u'''_2) \tau(E-\textfrac{3}{4}{u'''_2}^2)
F(u'''_2)\Bigg]~,
\label{eq:faddeev3}
\eea
where $\xi({\bf u_1},{\bf u_2})\equiv \langle{\bf u_1 u_2} | \xi\rangle$
 is defined as
\beq
\xi({\bf u_1},{\bf  u_2})=\exp\left(-\frac{u_1^2+\textfrac{3}{4}u_2^2}
   {\Lambda^2} \right)~.
\eeq
Note that the term in the exponent is the kinetic energy of
the three-body system. Thus, it is exactly the kinetic energy
flowing through the three-body interaction which is limited 
by the cutoff parameter $\Lambda$. This choice of the 
cutoff function  satisfies Bose symmetry explicitly.
The factors of $4\pi$ and $(4\pi)^3$ arise from our normalization 
of the S-wave projection of $t$ and $t_3$.

The value of the three-body
force is determined by the renormalization condition that  
the shallowest bound state energy is fixed as the cutoff is varied. 
This value is denoted by the horizontal solid
line in Fig.~\ref{fig:lambdab3}. Depending on the value of the cutoff,
the three-body force then must provide additional attraction or repulsion 
in order to keep the shallowest bound state energy fixed as the cutoff 
is varied.  Thus, one three-body datum (in our case the shallowest 
bound state energy) is required as input while all other low-energy
three-body observables can be predicted.
Once the shallowest bound state is fixed, the binding energies 
of the deeper bound states will also be cutoff independent. When the cutoff
is increased and a new bound state appears at threshold, the three-body
force must turn from strongly repulsive to strongly attractive to satisfy
the renormalization condition for the shallowest bound state. The 
additional state is then added as a deep state rather than at threshold.
Low-energy three-body observables are not affected by the additional deep 
bound states. As a consequence, the cutoff can be made arbitrarily 
large in the three-body system.

The renormalization procedure determines three-body coupling constant
$\lambda_3(B_3,\Lambda)$ uniquely. It was used in 
Refs.~\cite{Bedaque:1998kg,Bedaque:1998km} to renormalize the three-body
equation derived from nonrelativistic effective field theory
with an auxiliary field for the interacting two-particle state.
See Ref.~\cite{AFG95} for an earlier discussion of this renormalization
method. While the part of Eq.~(\ref{eq:faddeev3}) resulting from two-body
interactions only is very similar to the corresponding part of
the field-theoretical equation, the part containing the three-body force
(in form of $\tau_3$) is more complicated. The simplicity of the
field-theoretical equation is due the specific form of the three-body 
interaction using an auxiliary field in 
Refs.~\cite{Bedaque:1998kg,Bedaque:1998km}.\footnote{One would expect that
using auxiliary fields would also simplify the four-body
equations. However, it is not obvious how to treat the (2)+(2) clusters in 
the intermediate state without introducing uncontrolled approximations.}

We now explicitly verify the renormalization of the three-body system 
as described above.
By varying the cutoff parameter $\Lambda$ and tuning 
the three-body coupling $\lambda_3$ such that the shallowest three-body
binding energy stays constant, we can determine the renormalization
group evolution of $\lambda_3$ numerically. For the dimensionless coupling
constant $\lambda_3 \Lambda^4$, we confirm the results of
Refs.~\cite{Bedaque:1998kg,Bedaque:1998km,Wilson00,Mohr:2003du}:
$\Lambda^4\lambda_3(\Lambda)$ shows a limit cycle behavior and 
is single-valued.\footnote{Note that $\lambda_3$ is defined with
the opposite sign of the three-body coupling constant $H$ in the field 
theory formulation \cite{Bedaque:1998kg,Bedaque:1998km}.}
\begin{figure}[tb]
\centerline{\includegraphics*[width=5.in,angle=0]{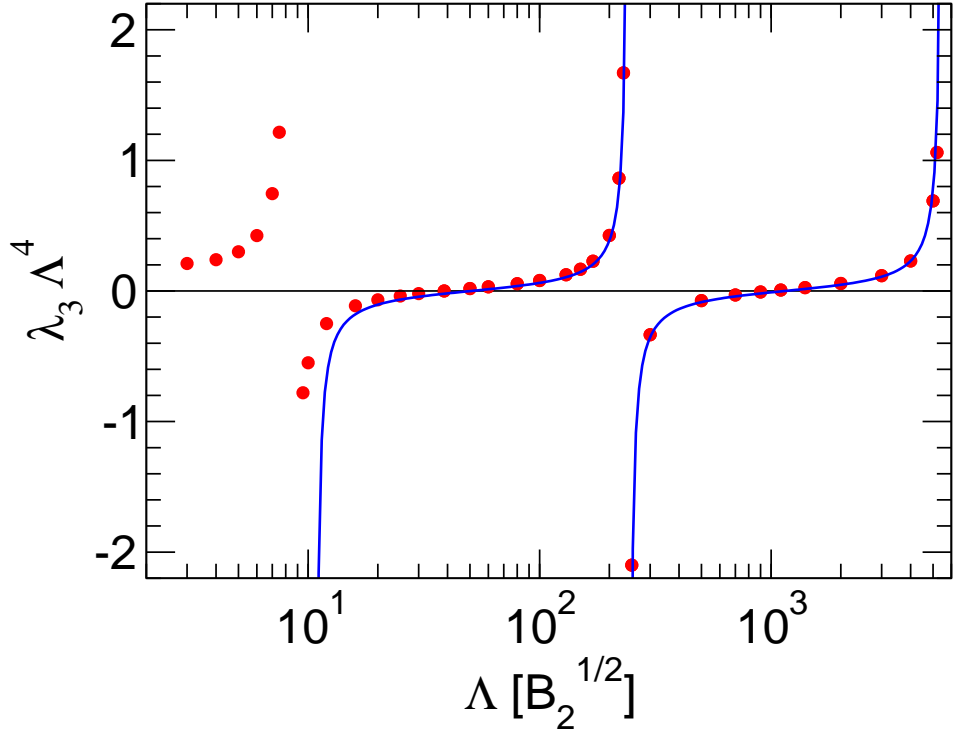}}
\caption{\label{fig:threeforce}The three-body coupling constant 
 $\lambda_3$ as a function of the cutoff parameter $\Lambda$.
 The solid blue line shows a fit of Eq.~(\ref{eq:limcyc}) with
 $c=0.076$ and $L_3 =23.3\sqrt{B_2}$ to the points for 
 $\Lambda \geq 245\sqrt{B_2}$.
}
\end{figure}
This limit cycle is shown in Fig.~\ref{fig:threeforce}. For large values
of the cutoff $\Lambda$, the dimensionless three-body coupling 
constant  $\Lambda^4\lambda_3$ flows towards an ultraviolet limit cycle.
For $\Lambda\to \infty$, it has the limiting behavior
\beq
\lambda_3(\Lambda)=\frac{c}{\Lambda^4} \;\frac{\sin(s_0 \ln(\Lambda/L_3)
  -\arctan(1/s_0))}{\sin(s_0 \ln(\Lambda/L_3)+\arctan(1/s_0))}~,
\label{eq:limcyc}
\eeq
where $s_0\approx 1.00624$ is a transcendental number that determines the
period of the limit cycle. If the cutoff $\Lambda$ is multiplied by 
a factor $\exp(n\pi/s_0) \approx (22.7)^n$ with $n$ an integer, the 
three-body coupling $\lambda_3$ is unchanged. 
$L_3$ is a three-body parameter generated
by dimensional transmutation. One can either specify the dimensionless
coupling constant $\Lambda^4 \lambda_3(\Lambda)$ and the cutoff $\Lambda$
or the dimensionful three-body parameter $L_3$. The constant 
$c$ is universal and independent of $L_3$ within our numerical accuracy.
We have determined $c$ numerically by fitting $\Lambda^4 \lambda_3(\Lambda)$
for different three-body parameters $L_3$ to Eq.(\ref{eq:limcyc}) and
 found:
\beq
c=0.074 \pm 0.003 \,,
\eeq 
where the error has been estimated from the observed variation in 
the fit results for $c$.
The three-body parameter $L_3$ can be determined by fixing
a three-body binding energy $B_3$.
Of course, one could also use a three-body binding energy directly
to characterize the value of the three-body coupling $\lambda_3$ at a 
given cutoff. However, it is advantageous to use $L_3$ because
Eq.~(\ref{eq:limcyc}) takes a particularly simple form in terms of $L_3$.

One might expect that the smooth Gaussian regulator function we use 
would lead to better behaved numerical solutions of the three-body
equations than the sharp cutoff used for the field-theoretical 
equation in Refs.~\cite{Bedaque:1998kg,Bedaque:1998km}.
However, it turns out that the convergence of the three-body coupling
to the ultraviolet limit cycle is significantly
slower than in the effective field theory formulation. 
This is due to the more complicated structure of the part of 
Eq.~(\ref{eq:faddeev3}) containing the three-body force. 
For very large loop momenta, the three-body part of 
Eq.~(\ref{eq:faddeev3}) simplifies considerably and has a similar form
as in the EFT formulation. In this limit, it is possible to derive the 
general form of Eq.~(\ref{eq:limcyc}).

In general, we recover the results for the three-body binding energies
from Refs.~\cite{Bedaque:1998kg,Bedaque:1998km,Wilson00,Mohr:2003du}.
However, we note that somewhat higher cutoffs are required to reach convergent 
results for the deeper three-body bound states. The numerical effort
for the solution of the three-body equations becomes larger as the 
value of the three-body coupling $\lambda_3$ is increased.
This is also related to the relatively complicated structure of the part of 
Eq.~(\ref{eq:faddeev3}) containing the three-body force. 

\subsection{The Four-Body Sector}
We now turn to the four-body sector. The four-body binding energies
are given by the non-trivial solutions of the Yakubovsky equations 
\cite{Yakubovsky:1966ue} which are based on
a generalization of the decomposition into Faddeev components 
for the three-body system. The full four-body wave function $\Psi$ is 
first decomposed into Faddeev components, followed by a second decomposition 
into so-called ``Yakubovsky components''. In the case of identical 
bosons, one ends up with two Yakubovsky components $\psi_A$ and
$\psi_B$. We start from the Yakubovsky equations
including a general three-body force in the form written down by
Gl\"ockle and Kamada \cite{Glockle:1993vr}. The full four-body 
bound state wave function is decomposed into the Yakubovsky
components $\psi_A$ and $\psi_B$ via
\beq
\label{eq:yak_wavefunction}
\Psi=(1+(1+P) P_{34})(1+P)\psi_A +(1+P)(1+\tilde{P})\psi_B~,
\eeq
where $P_{ij}$ exchanges particles $i$ and $j$, $P$ is defined in 
Eq.~(\ref{eq:defP}), and $\tilde{P}$ is given by
\beq
\tilde{P}=P_{13}P_{24}~.
\eeq
The equations for the two wave function components read:
\bea
\psi_A &=& G_0 t_{12} P [ (1+P_{34})\psi_A+\psi_B]
+\frac{1}{3}(1+G_0 t_{12})G_0 V_{3} \Psi~, \nonumber \\
\psi_B &=& G_0 t_{12} \tilde{P} [ (1+P_{34})\psi_A+\psi_B]~,
\label{eq:yaku}
\eea
where $t_{12}$ denotes the two-body t-matrix for particles 1 and 2
and $V_3$ is the three-body force defined in Eq.~(\ref{def:3pot}).
Note that the three-body force couples to the 
full four-body wave function $\Psi$. The factor of one third in front 
of the three-body force term arises because we insert the full 
three-body interaction for $V_3$. This is possible
since we consider three-body contact interactions which are symmetric 
under the exchange of any pair of particles.

In order to describe the four-body system at rest, three Jacobi
momenta are required. The structure of the four-body equations is more 
complex than that of the three-body equations because both (3)+(1) and 
(2)+(2) fragmentations can occur. As a consequence, two different sets
of Jacobi momenta are required. The (3)+(1) fragmentation is described
by the vector 
\beq
{\bf u_3}=\frac{3}{4}\bigl[{\bf k_4}-\frac{1}{3}(
{\bf k_1}+{\bf k_2}+{\bf k_3})\bigr]~,
\eeq
along with the Jacobi momenta for the three-body system 
given in Eq.~(\ref{eq:jacobi:3body}).
The $(2)+(2)$ fragmentation is described by the set:
\bea
{\bf v_1}&=&\frac{1}{2}({\bf k_1}-{\bf k_2})~,\quad
{\bf v_2}=\frac{1}{2}({\bf k_1}+{\bf k_2})
-\frac{1}{2}({\bf k_3}+{\bf k_4})~,\quad\mbox{and}\quad
{\bf v_3}=\frac{1}{2}({\bf k_3}-{\bf k_4})~.
\eea
Depending on which Yakubovsky component and operator in Eq.~(\ref{eq:yaku})
is concerned, one of the two sets of coordinates will be more convenient
than the other.

As in the previous subsection, we will only display the derivation of the
analytic expression for the Yakubovsky equations without the three-body 
force. It is natural to evaluate the Yakubovsky component $\psi_A$ as a
function of $u_1,u_2$, and $u_3$:
\bea
\langle u_1 u_2 u_3|\psi_A \rangle&=&4\pi\:
G_0(u_1, u_2,u_3)g(u_1)
\tau(E-\textstyle{\frac{3}{4}}u_2^2-\textstyle{\frac{2}{3}}u_3^2)
\nonumber\\[0.4cm]
&&\times\biggl[\int \hbox{d}u'_2\:{u'_2}^2
\int_{-1}^{1}\hbox{d}x\:g(\pi(u_2,u'_2))
\langle\pi(u'_2,u_2) u'_2 u_3|\psi_A\rangle\nonumber\\[0.4cm]
&&\quad+\frac{1}{2}\int\hbox{d}u'_2\:{u'_2}^2
\int_{-1}^1\hbox{d}x\int_{-1}^1\hbox{d}x'\:
g(\pi(u_2,u'_2))\nonumber\\
&&\qquad\qquad\qquad\qquad\qquad\times
\langle\pi(u'_2,u_2) \tilde{u}_2(u'_2,u_3) 
\tilde{u}_3(u'_2,u_3)|\psi_A\rangle\nonumber\\[0.4cm]
&&\quad+\frac{1}{2}\int\hbox{d}u'_2
\:{u'_2}^2
\int_{-1}^1\hbox{d}x\:\int_{-1}^1\hbox{d}x'\:
g(\pi(u_2,u'_2))\nonumber\\
&&\qquad\qquad\qquad\qquad\qquad\times
\langle\pi(u'_2,u_2) \tilde{v}_2(u'_2,u_3)
 \tilde{v}_3(u'_2,u_3)|\psi_B\rangle~\biggr]~,
\label{eq:psiA}
\eea
%
where the two-body propagator $\tau(E-\textstyle{\frac{3}{4}}u_2^2
-\textstyle{\frac{2}{3}}u_3^2)$ is again evaluated at the energy
of the corresponding two-body subsystem.
$G_0(u_1, u_2,u_3)$  is the free four-particle
propagator in the center-of-mass system of the four particles
and $\pi(u_2,u_2')$ is defined in Eq.~(\ref{eq:defpi}).
The second Yakubovsky component $\psi_B$ is computed as
a function of the momenta $v_1,v_2$, and $v_3$:
\bea
\langle v_1 v_2 v_3|\psi_B\rangle &=& 4\pi\: G_0(v_1,v_2,v_3)g(v_1)
\tau(E-\textstyle{\frac{1}{2}}v_2^2-v_3^3)
\bigg[\displaystyle{\int} 
\hbox{d}v'_3\:{v'_3}^2 g(v'_3)\langle v_3 v_2 v'_3|\psi_B\rangle
\nonumber\\[0.2cm]
&&\qquad\qquad+\int\hbox{d}v'_3\:
{v'_3}^2\int_{-1}^1\hbox{d}x\:g(v'_3)
\langle v_3 \:\bar{u}_2(v_2,v'_3)\: \bar{u}_3(v_2,v'_3)
|\psi_A\rangle \bigg] ~.
\label{eq:psiB}
\eea
In Eqs.~(\ref{eq:psiA}, \ref{eq:psiB}), we have used the 
abbreviations
\bea
\tilde{u}_2(u_2,u_3)&=&\sqrt{\textstyle{\frac{1}{9}}u_2^2+
\textstyle{\frac{64}{81}}u_3^2
+\textstyle{\frac{16}{27}}u_2 u_3 x'} 
~,\quad
\tilde{u}_3(u_2,u_3)=\sqrt{u_2^2+
\textstyle{\frac{1}{9}}u_3^2-\textstyle{\frac{2}{3}}u_2u_3x'}~,
\nonumber \\[0.3cm]
\tilde{v}_2(u_2,u_3)&=&\sqrt{u_2^2+\textstyle{\frac{4}{9}}u_3^2
+\textstyle{\frac{4}{3}}u_2 u_3 x'}
~,\quad
\tilde{v}_3(u_2,u_3)=\sqrt{\textstyle{\frac{1}{4}}u_2^2
+\textstyle{\frac{4}{9}u_3^2}
-\textstyle{\frac{2}{3}}u_2 u_3 x'}~,
\nonumber \\[0.3cm]
\bar{u}_2(v_2,v_3)&=&\sqrt{\textstyle{\frac{4}{9}}v_2^2
+\textstyle{\frac{4}{9}}v_3^2-\textstyle{\frac{8}{9}}v_2 v_3 x'}
~,\quad
\bar{u}_3(v_2,v_3)=\sqrt{\textstyle{\frac{1}{4}}v_2^2+v_3^2+v_2 v_3 x'}~.
\eea
Similar to the three-body case, 
Eqs.~(\ref{eq:psiA}, \ref{eq:psiB}) can be simplified
from two coupled integral equations in three variables to two 
coupled integral equations in two variables. Performing the substitutions
\bea
\langle u_1 u_2 u_3 |\psi_A\rangle&=&g(u_1)G_0(u_1,u_2,u_3)F_A(u_2,u_3)~,
\nonumber \\
\langle v_1 v_2 v_3 |\psi_B\rangle&=&g(v_1)G_0(v_1,v_2,v_3)F_B(v_2,v_3)~,
\eea
we obtain the integral equations for $F_A(u_2,u_3)$ and $F_B(v_2,v_3)$:
\bea
F_A(u_2,u_3)&=&4\pi\:
\tau(E-\textstyle{\frac{3}{4}}u_2^2-\textstyle{\frac{2}{3}}u_3^2)
\nonumber\\[0.3cm]
&&\times\biggl[\int \hbox{d}u'_2\:{u'_2}^2
\int_{-1}^{1}\hbox{d}x\:g(\pi(u_2,u'_2))
g(\pi(u'_2,u_2))\nonumber\\
&&\qquad\qquad\qquad\qquad\qquad\qquad\times
G_0(\pi(u'_2,u_2),u_2',u_3)
F_A (u_2',u_3)\nonumber\\[0.3cm]
&&\quad+\frac{1}{2}\int\hbox{d}u'_2\:{u'_2}^2
\int_{-1}^1\hbox{d}x\int_{-1}^1\hbox{d}x'
g(\pi(u_2,u'_2))g(\pi(u'_2,u_2))\nonumber\\
&&\qquad\quad\times
G_0(\pi(u'_2,u_2),\tilde{u}_2(u'_2,u_3),\tilde{u}_3(u'_2,u_3))
F_A(\tilde{u}_2(u'_2,u_3),\tilde{u}_3(u'_2,u_3))\nonumber\\[0.4cm]
&&\quad+\frac{1}{2}\int\hbox{d}u'_2\:{u'_2}^2
\int_{-1}^1\hbox{d}x\:\int_{-1}^1\hbox{d}x'\:g(\pi(u_2,u'_2))
g(\pi(u'_2,u_2))\nonumber\\
&&\qquad\quad\times
G_0(\pi(u'_2,u_2),\tilde{v}_2(u'_2,u_3),\tilde{v}_3(u'_2,u_3))
F_B(\tilde{v}_2(u'_2,u_3),\tilde{v}_3(u'_2,u_3))\biggr]~,
\\[0.3cm]
F_B(v_2,v_3)&=&4\pi\: \tau(E-\textstyle{\frac{1}{2}}v_2^2-v_3^2)
\biggl[\displaystyle{\int}\hbox{d}v'_3\:{v'_3}^2
g(v'_3)g(v_3)G_0(v_3,v_2,v'_3)F_B(v_2,v'_3)
\nonumber\\[0.3cm]
&&+\int\hbox{d}v'_3\:
{v'_3}^2\int_{-1}^1\hbox{d}x\: g(v'_3)g(v_3)
G_0(v_3,\bar{u}_2(v_2,v'_3),\bar{u}_3(v_2,v'_3))
\nonumber\\
&&\qquad\qquad\qquad\qquad\qquad\qquad\qquad\qquad\quad\times
F_A(\bar{u}_2(v_2,v'_3),\bar{u}_3(v_2,v'_3))\biggr]~.
\eea
The inclusion of the three-body force term is straightforward 
but lengthy and the corresponding expressions are given in the appendix.

In order to obtain the four-body binding energies, we have to solve
the Yakubovsky equations with the three-body force term. 
The binding energies can be found by discretizing the above equations 
and calculating the eigenvalues of the resulting matrix. 
They are given by the energies at which any eigenvalue of the matrix is 
equal to one. The wave function is then given by the corresponding 
eigenvector. 

The renormalization analysis of the four-body system is complicated 
by the cutoff dependence of the number of  bound states in the three-body 
subsystems. The further the cutoff $\Lambda$ is increased, the more
three-body bound states appear. While the spurious deep 
three-body states have no influence on low-energy 
three-body observables, they create an instability in the four-body system
which can collapse into a deep three-body bound state plus another particle.
This limits cutoff variations to an interval $\Lambda_0 < \Lambda < 
22.7 \,\Lambda_0$ for some $\Lambda_0$, in which the number of three-body 
bound states remains constant. Since the cutoff can still
be varied by more than a factor of ten, we are nevertheless 
able to study the renormalization properties and 
obtain converged numerical results. Alternatively, one
could explicitly subtract out the spurious bound states from
the three-body t-matrix. 
We will come back to this question in the next section.
\section{Renormalization and Numerical Results}
\label{sec:renres}
In this section, we will discuss the renormalization of the four-boson
system and present some numerical results for the four-body system of 
$^4$He atoms. For convenience, we will set Boltzmann's constant 
to unity: $k=1$.
Since the scattering length of $^4$He atoms is much larger than their 
effective range, they are an ideal application for our theory
and a leading order calculation should be accurate to about 10\%
(since $l/a\simeq 10\%$). 

The quantitative experimental information on low-energy $^4$He atoms, 
however, is rather limited.
Using diffraction of a molecular beam of small $^4$He clusters from a 
transmission grating, the bond length of the $^4$He dimer has
been measured to be $\langle r\rangle= (52 \pm 4)$ \AA\ \cite{Gri00}.
This value is an order magnitude larger than their effective range 
$r_e \approx 7$ \AA, which can be taken as an estimate of the
natural low-energy length scale $l$.
The scattering length $a=\left(104^{+8}_{-18}\right)$\AA\ and the
dimer binding energy $B_2=\left(1.1^{+0.3}_{-0.2}\right)$ mK
were derived from the measured bond length using the zero range 
approximation \cite{Gri00}. The $^4$He trimer, tetramer, and several larger 
$^4$He clusters have been observed \cite{STo96,BST02}, but
no quantitative experimental information about their binding energies
is available to date. 

However, there is a large number of theoretical calculations using realistic 
$^4$He potentials for the trimer ($^4$He$_3$). 
These calculations typically predict a trimer ground state with an 
energy of about 120 mK and one excited state with a binding energy of about
2 mK \cite{NFJ98,RoY00,MSSK01,Barl02}. 
The ground and excited states of the tetramer ($^4$He$_4$) and larger clusters 
have been calculated by Blume and Greene (BG) \cite{Blume:2000}. 
They have used the LM2M2 potential \cite{LM2M2}
and a combination of Monte Carlo methods and the 
adiabatic hyperspherical approximation. Their results for the trimer
energies agree with the exact three-body calculations of 
Refs.~\cite{NFJ98,RoY00,MSSK01,Barl02}. 

In the absence of quantitative experimental information on the three-body
clusters, we take the binding energy of the $^4$He trimer
excited state from theoretical calculations using the LM2M2 potential
as input to fix $L_3$. We use the value of BG: $B_3^{(1)}/B_2=1.767$
for this purpose \cite{Blume:2000,Lew97}. 
We can then calculate the trimer ground state and the tetramer binding 
energies for the LM2M2 potential based on low-energy universality. 
Before we present our numerical results for the $^4$He$_4$
system, we discuss the renormalization of the four-body problem. 

\begin{figure}[t]
\centerline{\includegraphics*[width=5.in,angle=0]{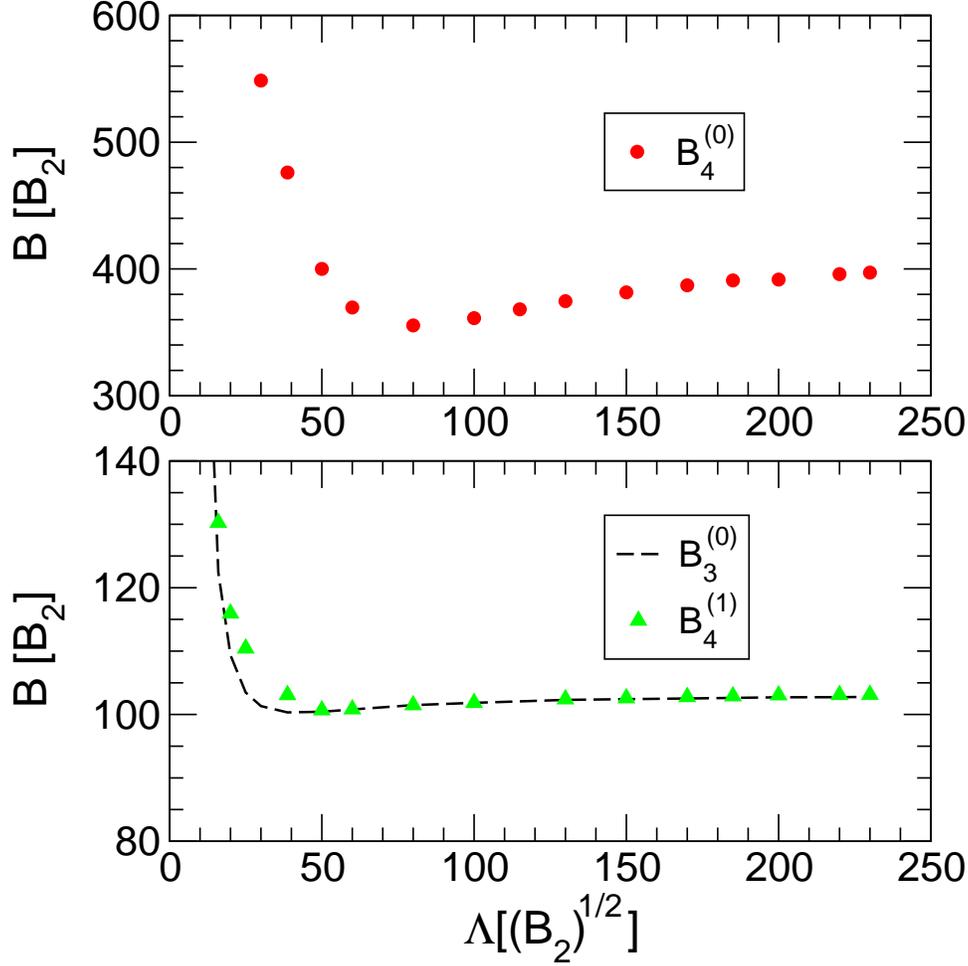}}
\caption{\label{fig_4bind}Binding energies of the three-
and four-body system as a function of the cutoff $\Lambda$.
$B_n^{(0)}$ and $B_n^{(1)}$ denote the ground and first excited state
of the $n$-body system.}
\end{figure}

In Fig.~\ref{fig_4bind}, we have plotted the three-body ground state
energy $B_3^{(0)}$ and the four-body energies as a function of the cutoff
$\Lambda$. As in the case of the trimer, the $^4$He tetramer 
has a ground state $B_4^{(0)}$ and one excited state $B_4^{(1)}$.
The cutoff dependence of $B_3^{(0)}$ must flatten out and
reach a plateau as $\Lambda$ is increased since the three-body system
was renormalized by fixing the trimer excited state binding energy
$B_3^{(1)}$  (cf. subsection \ref{sec:effT:3body}).
However, Fig.~\ref{fig_4bind} shows that the four-body binding energies 
$B_4^{(0)}$ and $B_4^{(1)}$ also reach a plateau as $\Lambda$ is increased.
The excited state energy $B_4^{(1)}$ has a negligible cutoff 
dependence already at fairly small cutoffs.
For the ground state $B_4^{(0)}$, the situation is somewhat more complicated 
and the cutoff dependence of $B_4^{(0)}$ reaches a plateau only at the largest 
cutoff values calculated.
The residual cutoff dependence is about 2\% for the excited state and
5\% for the ground state.
A slower convergence for the ground state is expected since the value for
$B_4^{(0)}$ is a factor four larger than for the excited state and finite 
cutoff effects of the order $\sqrt{|E|}/\Lambda$ are more important.
The residual cutoff dependence for both states is at least a factor 2
smaller than the corrections from higher orders in the expansion in $l/a$ 
which are expected to be of the order $l/a\sim 10\%$.
We speculate that higher precision could be achieved by increasing 
the cutoff further. As already noted earlier, this would create unphysical
three-body bound states which create an instability in the four-body
system. These states would have to be subtracted explicitly.
While such a subtraction is possible, this is beyond the scope of our 
paper and we will not attempt such a subtraction here.

Taken together, the above observations provide strong numerical evidence 
that the four-body binding energies are cutoff independent up to higher
order corrections in  $l/a$. In particular, a four-body force 
with limit cycle behavior would lead to a much stronger cutoff dependence
of the binding energies (cf.~Fig.~\ref{fig:lambdab3}) and can be excluded.
The occurrence of the plateaus for $B_4^{(0)}$ and $B_4^{(1)}$ in 
Fig.~\ref{fig_4bind} suggests that a four-body force is not required for 
renormalization of the four-body system at leading order in $l/a$.
Renormalization of the three-body system 
automatically generates cutoff-independent results for the four-body 
binding energies. 
As a consequence, the four-body binding energies can be predicted 
from two- and three-body input alone.

We now turn to our numerical results for the four-body system of
$^4$He atoms.
From the plateaus in Fig.~\ref{fig_4bind}, we can read off the values of 
the binding energies.
A comparison of our results with the values obtained by BG
\cite{Blume:2000} is shown in Table \ref{tab:results}.
The results of their calculation for the trimer and tetramer are given in the 
two right columns of Table~\ref{tab:results}, while our results are given in 
the two left columns.
\begin{table}[t]
\begin{center}
\begin{tabular}{|c||c|c||c|c|}
\hline
system & $B^{(0)}$ [mK] & $B^{(1)}$ [mK] & $B_{\rm BG}^{(0)}$ [mK] 
& $B_{\rm BG}^{(1)}$ [mK] \\ \hline\hline
$^4$He$_3$ & 127 & [2.186]   & 125.5  & 2.186\\
$^4$He$_4$ & 492 & 128  & 559.7  & 132.7\\ \hline
\end{tabular}
\end{center}
\caption{\label{tab:results}Binding energies of the $^4$He
trimer and tetramer in mK.
The two right columns show the results by Blume and Greene 
\cite{Blume:2000} (denoted by the index BG)
while the two left columns show our results. The number in brackets was
used as input to fix $L_3$.}
\end{table}
In general, our results are in good agreement with the values of BG.
For the trimer ground state and the tetramer excited state
energies, we obtain the values $B_3^{(0)}=127$ mK and $B_4^{(1)}=
128$ mK, respectively. For the tetramer ground state, 
we obtain $B_4^{(0)}=492$ mK. While the value of $B_4^{(0)}$ is 
already relatively large, it is still a factor three smaller than the 
natural four-body energy scale $\sim 1.5$ K where the effective theory 
description is expected to break down.

The natural energy scales can be estimated as follows:
For two particles, it is directly determined by the natural length scale $l$ 
and the mass $M$ of the particles: $e_2\sim\hbar^2/(M\,l^2)\approx 250$ mK. 
For three and four particles, this estimate should be scaled according
to the number of pairs available,\footnote{We are grateful
to Eric Braaten for suggesting this scaling of the natural
energy scale according to the number of pairs to us.} leading to the values 
$e_3\sim 750$ mK and $e_4\sim 1.5$ K. 
This estimate can be made for cutoff values at which
the three-body force vanishes. Since all observables are independent
of the cutoff, however, it is valid for arbitrary cutoffs.
Our short-range effective theory can describe $n$-body bound 
states with binding energies $B_n \ll e_n$. For deeper bound states
closer to the natural energy scale the errors are expected to increase.

The values in Table \ref{tab:results} 
have been computed at a cutoff of $\Lambda=235 \sqrt{B_2}$
which is close to the largest possible value with only two 
three-body bound states. Our values for $B_4^{(0)}$ and $B_4^{(1)}$
agree with the BG values to within 12\% and 3\%, respectively.
The dominant correction to our results is due to effective range 
effects which are not included in our leading order calculation. 
These deviations are within the expected accuracy of the effective theory.
We expect the effective range corrections to the leading order result
to be of the order $r_e/a \sim 10\%$.
From the residual cutoff dependence, we estimate the numerical error of 
our calculation to be of the order of 2\% for the excited state and
5\% for the ground state. For fixed value of the cutoff $\Lambda$,
our calculations are numerically accurate to 3 significant digits.

The large scattering length of $^4$He atoms also leads to universal
properties in the four-body problem. A typical example is 
the existence of correlations between different observables. These
correlations become manifest in universal scaling functions
relating dimensionless combinations of observables. Various scaling
functions for the three-body system of $^4$He atoms were studied in 
Refs.~\cite{Braaten:2002jv,FTD99,DFT00}. 
Here, we consider the four-body binding energies as a function of the
three-body binding energies. 
In phenomenological calculations of the four-nucleon system,
an approximately linear correlation between the three- and four-nucleon 
binding energies for various nucleon-nucleon potentials is observed:
the Tjon line \cite{Tjo75}. This correlation is approximately linear
for the relevant range of binding energies \cite{Nogga:2000uu}.
Since the $^4$He trimer and tetramer have both a ground state and
an excited state, there are four \lq\lq Tjon lines'' in this case.

\begin{figure}
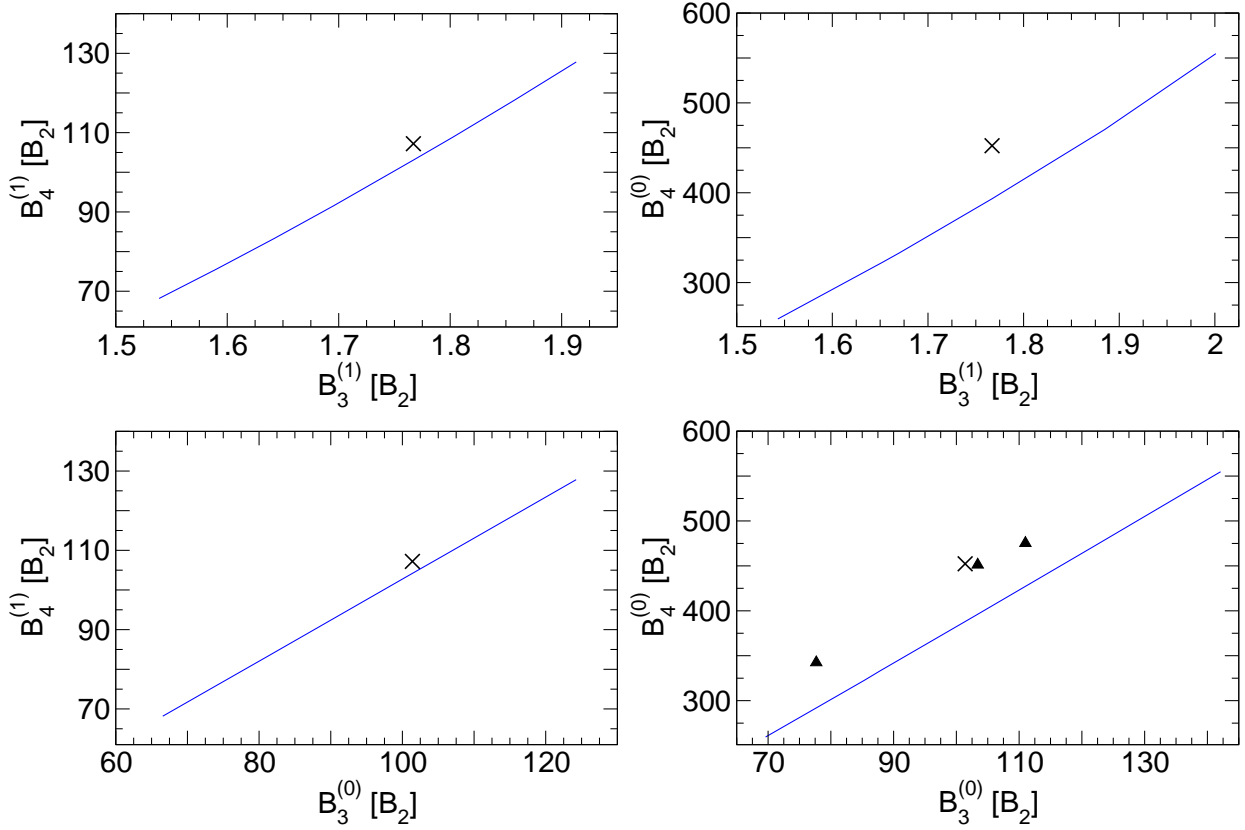

\centerline{\includegraphics*[width=3.2in,angle=0]{tjon_41.eps}
\includegraphics*[width=3.2in,angle=0]{tjon_40.eps}}
\caption{\label{fig:tjonline}
The correlations between
the ground and excited state energies of the $^4$He trimer and tetramer.
Upper row: the four-body excited state energy $B_4^{(1)}$ (left panel)
and ground state energy $B_4^{(0)}$ (right panel) 
as a function of the three-body excited state energy $B_3^{(1)}$.
Lower row: the same quantities as a function of three-body ground
state energy $B_3^{(0)}$. The solid line shows the leading order 
effective theory result and the cross denotes the calculation for
the LM2M2 potential by Blume and Greene \cite{Blume:2000}. The triangles
show the results for the TTY, HFD-B, and HFDHE2 potentials \cite{Lew97,Naka83}.
}
\end{figure}

In Fig.~\ref{fig:tjonline}, we show the correlations between
the ground and excited state energies of the $^4$He trimer and tetramer.
The two plots in the upper row show the energies of the tetramer excited
state $B_4^{(1)}$ (left panel) and ground state $B_4^{(0)}$ (right panel) 
as a function of trimer excited state energy $B_3^{(1)}$, while the 
lower row shows the same quantities as a function of the trimer
ground state energy $B_3^{(0)}$. The solid line is the leading order 
result of our effective theory calculation and the cross denotes the result
of the calculation by BG for the LM2M2 potential \cite{Blume:2000}. 
For the ground states of the trimer and tetramer, calculations with
other $^4$He potentials are available as well. As an example, we show 
the results for the TTY, HFD-B, and HFDHE2 potentials taken from 
Refs.~\cite{Lew97,Naka83}.

Similar to the  nuclear sector, we find an approximately linear
correlation over the range of binding energies relevant to $^4$He atoms.
The calculations for the realistic $^4$He potentials fall close to the 
universal scaling curves from our effective theory. For the correlation
between the ground state energies, the \lq\lq Tjon line'' is directly evident
in the potential model calculations shown in Fig.~\ref{fig:tjonline}. 
If calculations with other 
potentials were performed for the excited state energies,
they would also fall on a line parallel to the universal scaling curve from
effective theory. The deviation of calculations
using realistic potentials from the universal line is mainly governed by 
effective range corrections which are expected to be of the order
10\% for $^4$He atoms. For the tetramer ground state, this deviation is 
about a factor four larger than for the excited state.

While the correlations in Fig.~\ref{fig:tjonline} are approximately linear
over the range of binding energies calculated, we expect them to become
nonlinear for a sufficiently large range of binding energies.
Similar nonlinearities were observed for correlations between
three-body observables in Refs.~\cite{Bedaque:1998km,Braaten:2002jv}.
Our results suggest that the \lq\lq Tjon lines'' are universal properties
of few-body systems with short-range interactions and large scattering 
length. They do not depend on the details of the short-distance 
physics which are very different in atomic and nuclear systems.
It is interesting to note that in nuclear physics only a correlation 
between the ground-state energies of the three- and four-body 
systems has been observed. We expect this correlation to hold for
all S-wave states that are within the range of validity of an
effective theory with contact interactions. It would be interesting to 
see whether such a correlation also holds for the excited $J^P=0^+$
state of the $\alpha$ particle above the $p+^3$H threshold.

We have fitted the scaling functions shown in Fig.~\ref{fig:tjonline}
with linear expressions and obtained:
\bea
\label{eq:tjon1}
\frac{B_4^{(0)}}{B_2} &=& -24.752 +4.075\;\frac{B_3^{(0)}}{B_2}~,
\qquad\qquad 69 \leq  \frac{B_3^{(0)}}{B_2} \leq 142~,
\\ \label{eq:tjon2}
\frac{B_4^{(0)}}{B_2} &=& -742.0 +645.1\;\frac{B_3^{(1)}}{B_2}~,
\qquad\qquad  1.54 \leq \frac{B_3^{(1)}}{B_2} \leq 2.00~,
\\ \label{eq:tjon3}
\frac{B_4^{(1)}}{B_2} &=& -0.662 + 1.034 \;\frac{B_3^{(0)}}{B_2}~,
\qquad\qquad 65 \leq  \frac{B_3^{(0)}}{B_2} \leq 125~,
\\
\frac{B_4^{(1)}}{B_2} &=& -178.0 + 159.4 \;\frac{B_3^{(1)}}{B_2}~,
\qquad \qquad
 1.52 \leq \frac{B_3^{(1)}}{B_2} \leq 1.92~.
\label{eq:tjon4}
\eea
These relations can be used to predict the tetramer ground
and excited state energies for any potential for which
one of the trimer energies and the dimer binding energy are known.
The expressions (\ref{eq:tjon1})-(\ref{eq:tjon4}) 
are of the same accuracy as our explicit calculations (see the 
discussion above). They are expected to be most accurate
for the excited states.
\section{Summary and Outlook}
\label{sec:conc}
In this paper, we have studied the four-body system with short-range
interactions and large scattering length. We have 
concentrated on the bound state problem of four bosons 
starting from the Yakubovsky equations \cite{Yakubovsky:1966ue}.
We have constructed an effective interaction potential including 
both a two- and three-body contact interaction.
This is the minimal set of contact interactions required for 
renormalization of the three-body problem \cite{Bedaque:1998kg,Bedaque:1998km}.
The two parameters of the effective potential were determined from
matching to the binding energy of the dimer and the excited state
of the trimer. We have then solved the four-body bound state problem
under the assumption that no four-body interaction is required for
renormalization at leading order. We found that
after renormalizing the two- and three-body subsystems, the four-body binding 
energies were automatically independent of the ultraviolet
cutoff. This result suggests 
that the four-body interaction is not of leading order and the low-energy
four-body observables are determined by properties of the two- and 
three-body systems up to corrections suppressed by $l/|a|$.
Although we have considered only the four-boson bound state problem
with large scattering length explicitly, we expect this result to hold for
all low-energy four-body observables.

We have applied this effective theory to the four-body system of
$^4$He atoms and calculated the ground and excited state energies
of the $^4$He tetramer. In the absence of experimental information
on the $^4$He trimer, we have taken the excited state energy of the 
$^4$He trimer as calculated by Blume and Greene for the LM2M2
potential \cite{Blume:2000} as input to determine the three-body
parameter $L_3$. For the binding energies of the trimer ground state and 
the tetramer ground and excited states, we find the values
$B_3^{(0)}=127$ mK, $B_4^{(1)}= 128$ mK, and  $B_4^{(0)}=492$ mK,
respectively. The latter energy is still about a factor three
smaller than the natural four-body energy scale $\sim 1.5$ K where
the effective theory is expected to break down.
Our values for $B_4^{(0)}$ and $B_4^{(1)}$
agree with the calculation by Blume and Greene to within 12\% and 3\%, 
respectively. These deviations are consistent with the expected accuracy 
at leading order in the large scattering length of
about $l/|a|\sim 10\%$.

The large scattering length of $^4$He atoms leads to universal
properties such as universal scaling functions.
We have calculated the universal scaling functions relating 
the tetramer energies to the trimer energies.
The correlations are approximately linear in the region of
binding energies relevant for $^4$He atoms. 
As expected from low-energy universality,
the results of various calculations using realistic $^4$He potentials 
fall close to the universal scaling curves. 
Corrections to the scaling curves are mainly governed by effective range 
effects. We have fitted the calculated scaling functions with 
linear expressions (\ref{eq:tjon1})-(\ref{eq:tjon4}) that
can be used to obtain the tetramer binding energies at leading order 
in $l/|a|$ for any potential if one of the trimer
binding energies is known.

There are a number of directions that should be pursued in future
work. While we have demonstrated that a four-body force is not
necessary to renormalize the four-body system to leading order,
the general power counting for four-body forces is still not understood.
At which order does the leading four-body interaction enter? In the 
three-body system, e.g., the first order correction is due to the two-body 
effective range. If a similar situation holds in the four-body system, 
it would be possible to predict low-energy four-body observables up to 
corrections of order $(l/a)^2$ from two- and three-body information alone. 

The extension of the effective theory to calculate four-body scattering
observables would be very valuable. 
The knowledge of the dimer-dimer scattering length, for example,
is important for experiments with ultracold atoms. For the simpler
problem of fermions with two spin states (where the three-body parameter
$L_3$ does not contribute), the dimer-dimer scattering length
was recently calculated \cite{petrov}.

Whether this effective theory can be applied 
to the nuclear four-body system like the pionful theory
\cite{Epelbaum:2002vt} is an open question.
While it is straightforward to generalize the effective theory to 
include spin and isospin, it is not clear whether an effective theory
without explicit pions will be adequate for the $\alpha$ particle
ground state
with a binding energy of about 28 MeV. This question deserves further
study. The effective theory might also help to shed some light on the
renewed speculations about the existence of a shallow 
tetraneutron bound state \cite{4neutron}.

\begin{acknowledgments}
We thank E.~Braaten, W.~Gl{\"o}ckle, U. van Kolck, and A.~Nogga 
for valuable discussions and 
E.~Braaten and A.~Nogga for comments on the manuscript.
This research was supported in part by DOE grant DE-FG02-00ER41132.
\end{acknowledgments}
\begin{appendix}
\section{Yakubovsky Equations with Three-Body Force}
In this appendix, we derive the analytic expressions 
for the three-body force term in the  Yakubovsky equations (\ref{eq:yaku})
in momentum space. The three-body force term
\beq
\label{eq:app_threebodyterm}
\frac{1}{3}\langle u_1 u_2 u_3|(1+G_0 t_{12})G_0 V_{3}|\Psi\rangle~,
\eeq
couples to the full four-body wave function $\Psi$ 
(cf.~Eq.~(\ref{eq:yak_wavefunction})), which
is related to the Yakuvosky components $\psi_A$ and $\psi_B$ via
\bea
\Psi &=& (1+P_{34}+PP_{34}+P+P_{34}P+PP_{34}P)\psi_A+
(1+P+\tilde{P}+P\tilde{P})\psi_B~.
\eea

In order to simplify Eq.~(\ref{eq:app_threebodyterm}), we consider a term
with an arbitrary permutation operator $X$ acting on $\psi_A$.
Inserting the $S$-wave projection operator, Eq.~(\ref{eq:S-proj}),
we can write
\bea
&&\frac{1}{3}\langle u_1 u_2 u_3|(1+G_0 t_{12})G_0 V_{3} X
|\psi_A\rangle=
\nonumber\\
&&=\frac{\lambda_3}{3}(4\pi)^2 G_0(u_1, u_2, u_3)
\biggl[\int\hbox{d}u'_1 {u'_1}^2
\hbox{d}u'_2 {u'_2}^2\mathcal{D}u'' {u''}^2 \xi(u_1,u_2)
\xi(u'_1,u'_2)\nonumber\\
&&\hspace{5cm}\times\langle u'_1 u'_2 u_3|X|u''_1 u''_2 u''_3\rangle
\langle u''_1 u''_2 u''_3|\psi_A\rangle\nonumber\\[0.3cm]
&&\qquad+4\pi\int\hbox{d}u'_1 {u'_1}^2\hbox{d}u''_1 {u''_1}^2
\hbox{d}u''_2 {u''_2}^2 \mathcal{D}u''' {u'''}^2 g(u_1)
\tau(E-\textstyle{\frac{3}{4}}u_2^2-
\textstyle{\frac{2}{3}}u_3^2)g(u'_1)
\nonumber \\
&&\qquad\quad\times G_0(u'_1, u_2,u_3)\xi(u'_1,u_2)
\xi(u''_1,u''_2)\langle u''_1 u''_2 u_3|X|u'''_1 u'''_2 u'''_3\rangle
\langle u'''_1 u'''_2 u'''_3|\psi_A\rangle\biggr]~,
\eea
where the factors of $4\pi$ and $(4\pi)^2$ arise from the $S$-wave
projection of the two-body t-matrix and three-body potential, respectively.
This expression can be rewritten as:
\bea
&&\frac{1}{3}\langle u_1 u_2 u_3|(1+G_0 t_{12})G_0 V_{3} X
|\psi_A\rangle=
\nonumber\\
&&=\frac{\lambda_3}{3}(4\pi)^2 G_0(u_1,u_2,u_3)\Bigr[\xi(u_1,u_2)
+4\pi g(u_1)
\tau(E-\textstyle{\frac{3}{4}}u_2^2-\textstyle{\frac{2}{3}}u_3^2)
\mathcal{I}(u_2,u_3)\Bigr]\nonumber\\
&&\qquad\times\int\hbox{d}u'_1 {u'_1}^2 \hbox{d}u'_2 {u'_2}^2
\mathcal{D}u''{u''}^2 \xi(u'_1,u'_2)\langle u'_1 u'_2 u_3|
X|u''_1 u''_2 u''_3\rangle\langle u''_1 u''_2 u''_3|
\psi_A\rangle\nonumber\\
&&=\frac{\lambda_3}{3}(4\pi)^2 G_0(u_1,u_2,u_3)
\Bigr[\xi(u_1,u_2)+4\pi g(u_1)
\tau(E-\textstyle{\frac{3}{4}}u_2^2-\textstyle{\frac{2}{3}}u_3^2)
\mathcal{I}(u_2,u_3)\Bigr]K^{(A)}_X(u_3)~,
\eea
where we have defined the quantities
\beq
\mathcal{I}(u_1,u_2)=\int \hbox{d}u'_1 {u'_1}^2 G_0(u'_1,u_2,u_3)
g(u'_1) \xi(u'_1,u_2)~,
\eeq
and 
\beq
K_X^{(A)}(u_3)=\int\hbox{d}u'_1 {u'_1}^2 \hbox{d}u'_2 {u'_2}^2
\mathcal{D}u''{u''}^2 \xi(u'_1,u'_2)\langle u'_1 u'_2 u_3|
X|u''_1 u''_2 u''_3\rangle\langle u''_1 u''_2 u''_3|
\psi_A\rangle~.
\eeq
An analogous expression $K_X^{(B)}$ can be derived for permutation
operators which act on the second Yakubovsky component $\psi_B$.

We proceed by giving the analytical expressions for
the $K_X^{(A)}$ and $K_X^{(B)}$ which appear in the computation of the
three-body force term. There are six combinations of permutation
operators acting on $\psi_A$:
\begin{subequations}
\bea
K^{(A)}_{1}&=&\int\hbox{d}u'_1 {u'_1}^2\hbox{d}u'_2 {u'_2}^2
\xi(u'_1,u'_2)\langle u'_1 u'_2 u_3|\psi_A\rangle~,\\
K^{(A)}_{P_{34}}&=&\frac{1}{2}\int\hbox{d}u'_1 {u'_1}^2\hbox{d}u'_2
{u'_2}^2
\xi(u'_1,u'_2)\int_{-1}^1\hbox{d} x'\:\langle u'_1 
\tilde{u}_2(u'_2,u_3) \tilde{u}_3(u'_2,u_3)|\psi_A\rangle\\
K^{(A)}_{P P_{34}}&=&\frac{1}{2}\int \hbox{d} u'_2 {u'_2}^2
\hbox{d} u''_2 {u''_2}^2\int_{-1}^1\hbox{d} x
\int_{-1}^{1}\hbox{d}x'\: \xi(\pi(u'_2,u''_2),u'_2)
\nonumber\\
&&\qquad\qquad\qquad\qquad\qquad\times
\langle \pi(u''_2,u'_2) \tilde{u}_2(u''_2,u_3)
\tilde{u}_3(u''_2,u_3)|\psi_A\rangle~,\\
K^{(A)}_{P}&=&\int\hbox{d}u'_2{u'_2}^2 \hbox{d}u''_2
{u''_2}^2 \int^1_{-1}\hbox{d}x\:
\xi(\pi(u'_2,u''_2),u'_2)\langle \pi(u''_2,u'_2)
u''_2 u_3|\psi_A\rangle~,\\
K^{(A)}_{P_{34}P}&=&\frac{1}{2}\int\hbox{d}u'_2 {u'_2}^2
\hbox{d}u''_2 {u''_2}^2\int_{-1}^1\hbox{d}x\int_{-1}^1
\hbox{d}x'\: \xi(\pi(\tilde{u}_2(u'_2,u_3),u''_2),u'_2)
\nonumber\\
&&\qquad\qquad\qquad\times
\langle \pi(u''_2,\tilde{u}_2(u'_2,u_3))\: u''_2
\tilde{u}_3(u'_2,u_3|\psi_A\rangle~,\\
K^{(A)}_{P P_{34} P}&=&\frac{1}{2}\int\hbox{d}u'_2{u'_2}^2
\hbox{d}u''_2 {u''_2}^2 \int_{-1}^1\hbox{d}x \int_{-1}^1\hbox{d}x'
\int_{-1}^1\hbox{d}x''\:\xi(\pi(u'_2,u''_2),u'_2)\nonumber\\
&&\times\langle \hat{u}_1[\pi(u''_2,u'_2),
\tilde{u}_2(u''_2,u_3)]\hat{u}_2[\pi(u''_2,u'_2),
\tilde{u}_2(u''_2,u_3)]\tilde{u}_3(u''_2,u_3)|\psi_A\rangle~,
\eea
\end{subequations}
and four combinations of operators acting on $\psi_B$:
\begin{subequations}
\bea
K_{1}^{(B)}&=&\frac{1}{2}\int\hbox{d}u'_1 {u'_1}^2\hbox{d} u'_2 {u'_2}^2
\int_{-1}^1\hbox{d}x'\: \xi(u'_1,u'_2)\langle u'_1 \tilde{v}_2(u'_2,u_2)
\tilde{v}_3(u'_2,u_3)|\psi_B\rangle~,\\
K_{P}^{(B)}&=&\frac{1}{2}\int\hbox{d}u'_2 {u'_2}^2\hbox{d}u''_2
{u''_2}^2 \int_{-1}^1\hbox{d}x\int_{-1}^1\hbox{d}x'\:
\xi(\pi(u'_2,u''_2),u'_2)\nonumber\\
&&\qquad\qquad\qquad\qquad\qquad\times\langle \pi(u''_2,u'_2)
\tilde{v}_2(u''_2,u_3) \tilde{v}_3(u''_2,u_3)|\psi_B\rangle~,\\
K_{\tilde{P}}^{(B)}&=&\frac{1}{2}\int\hbox{d}u'_1{u'_1}^2
\hbox{d}u'_2{u'_2}^2\int_{-1}^1\hbox{d}x' \:\xi(u'_1,u'_2)
\langle \tilde{v}_3(u'_2,u_3) \tilde{v}_2(u'_2,u_3) u'_1|
\psi_B\rangle~, \\
K_{P\tilde{P}}^{(B)}&=&\frac{1}{2}\int\hbox{d}u'_2 {u'_2}^2
\hbox{d}u''_2{u''_2}^2\int_{-1}^1\hbox{d}x\int_{-1}^1
\hbox{d} x'\: \xi(\pi(u'_2,u''_2),u'_2)\nonumber\\
&&\qquad\qquad\qquad\qquad\qquad\times
\langle\tilde{v}_3(u''_2,u_3) \tilde{v}_2(u''_2,u_3)
\pi(u''_2,u'_2)|\psi_B\rangle~,
\eea
\end{subequations}
where $\hat{u}_1(u_1,u_2)$ and $\hat{u}_2(u_1,u_2)$ are defined
as
\beq
\hat{u}_1(u_1,u_2)=\sqrt{\textstyle{\frac{1}{4}}u_1^2
+\textstyle{\frac{9}{16}}u_3^3+\textstyle{\frac{3}{4}}u_1u_2x''}~, 
\quad \hat{u}_2(u_1,u_2)=\sqrt{u_1^2
+\textstyle{\frac{1}{4}}u_2^2-u_1u_2 x''}~.
\eeq
\end{appendix}

\end{document}